\documentclass[11pt,letterpaper]{article}

\usepackage{times}
\usepackage{relsize}
\usepackage{microtype}
\usepackage{cite}
\usepackage{xspace}
\usepackage[table,xcdraw]{xcolor}
\usepackage{color,colortbl}
\usepackage{subfigure}
\usepackage{alltt}
\usepackage{url}
\usepackage{comment}
\usepackage{graphicx}
\usepackage{relsize}
\usepackage{paralist}
\usepackage{todonotes}
\usepackage{wrapfig}
\usepackage{titlesec}
\usepackage{epsfig}
\usepackage{geometry}
\usepackage{tikz}
\usepackage[normalem]{ulem}
\usepackage{titlesec}
\usepackage{booktabs}
\usepackage{wrapfig}
\usepackage{multirow}
\usepackage{pifont}
\usepackage{wrapfig}
\usepackage{soul}
\usepackage{enumitem}
\usepackage{tcolorbox}
\usepackage{pdfpages}
\usepackage{fancyhdr}
\usepackage{longtable}
\usepackage{makecell}
\usepackage{lettrine}
\usepackage{listings}

\titlespacing{\section}{0pc}{1pc}{1pc}
\titlespacing{\subsection}{0pc}{2pc}{1pc}
\titlespacing{\subsubsection}{0pc}{1pc}{1pc}
\titleformat*{\section}{\Large\bfseries}
\titleformat*{\subsection}{\large\bfseries}
\titleformat*{\subsubsection}{\normalsize\bfseries}

\lstdefinestyle{wcsStyle}{
  tabsize=4,
  showspaces=false,
  showstringspaces=false,
  aboveskip=0em,
  belowskip=0em,
}

\definecolor{Gray}{gray}{0.9}

\newcommand{\pp}[1]{\medskip \noindent \textbf{\emph{#1.}}\xspace}

\begin{document}

\includepdf[pages=-]{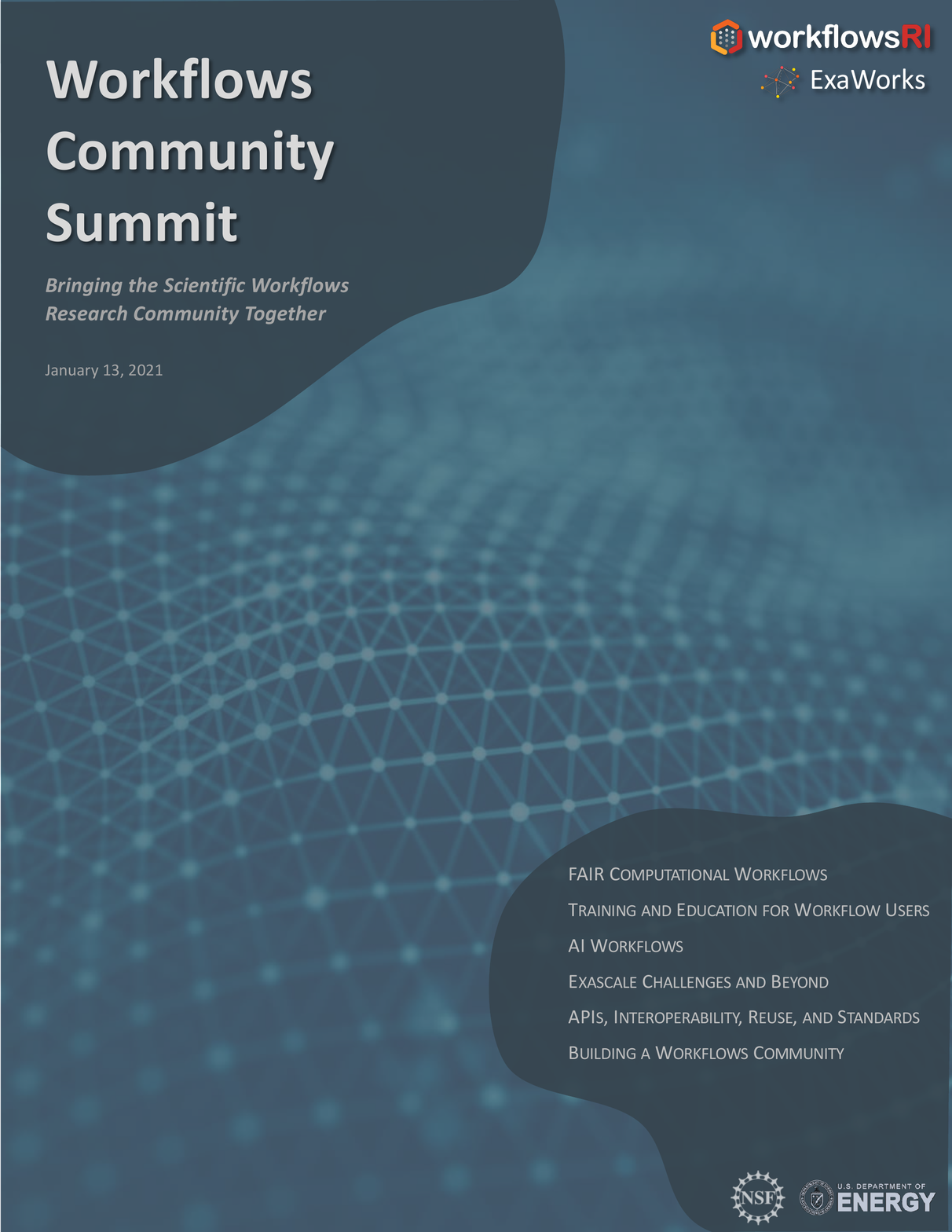}


\pagestyle{fancy}
\fancyhf{}
\rhead{
  \includegraphics[height=12pt]{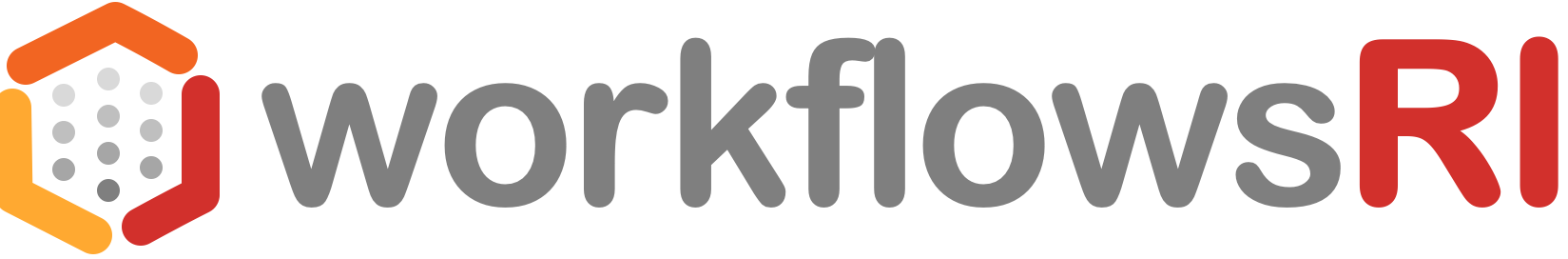} 
  \includegraphics[height=11pt]{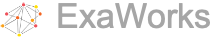}
}
\lhead{2021 Workflows Community Summit}
\rfoot{\thepage}


\begin{table}
\centering
\smaller
\vspace{1em}
\begin{tabular}{p{16cm}}
    \textbf{Disclaimer}
    \\
    The Workflows Community Summit was supported by the National Science Foundation (NSF) under grants number 2016610, 2016619, and 2016682, and the Department of Energy (DOE). Any opinions, findings, and conclusions or recommendations expressed at the event or in this report are those of the authors and do not necessarily reflect the views of NSF or DOE.
    \\
    \vspace{0.5em}
    \textbf{Preferred citation} 
    \\
    R. Ferreira da Silva, H. Casanova, K. Chard, D. Laney, D. Ahn, S. Jha, C. Goble, 
    L. Ramakrishnan, L. Peterson, B. Enders, D. Thain, I. Altintas, Y. Babuji, R. Badia, 
    V. Bonazzi, T. Coleman, M. Crusoe, E. Deelman, F. Di Natale, P. Di Tommaso, T. Fahringer,
    R. Filgueira, G. Fursin, A. Ganose, B. Gr\"uning, D. S. Katz, O. Kuchar, 
    A. Kupresanin, B. Ludascher, K. Maheshwari, M. Mattoso, K. Mehta, T. Munson, J. Ozik, 
    T. Peterka, L. Pottier, T. Randles, S. Soiland-Reyes, B. Tovar, 
    M. Turilli, T. Uram, K. Vahi, M. Wilde, M. Wolf, J. Wozniak, 
    ``Workflows Community Summit: Bringing the Scientific Workflows Research
    Community Together", Technical Report, March 2021, DOI: 10.5281/zenodo.4606958.
    \\
    \rowcolor[HTML]{F7F7F7}
    \lstset{basicstyle=\scriptsize,style=wcsStyle}
    \begin{lstlisting}
@misc{wcs2021community,
  author    = {Ferreira da Silva, R. and Casanova, H. and Chard, K. and 
               Laney, D. and Ahn, D. and Jha, S. and Goble, C. and 
               Ramakrishnan, L. and Peterson, L. and Enders, B. and Thain, 
               D. and Altintas, I. and Babuji, Y. and Badia, R. and Bonazzi, 
               V. and Coleman, T. and Crusoe, M. and Deelman, E. and Di 
               Natale, F. and Di Tommaso, P. and Fahringer, T. and Filgueira,
               R. and Fursin, G. and Ganose, A. and Gruning, B. and  
               Katz, D. S. and Kuchar, O. and Kupresanin, A. and 
               Ludascher, B. and Maheshwari, K. and Mattoso, M. and Mehta, 
               K. and Munson, T. and Ozik, J. and Peterka, T. and Pottier, L. 
               and Randles, T. and Soiland-Reyes, S. and
               Tovar, B. and Turilli, M. and Uram, T. and 
               Vahi, K. and Wilde, M. and Wolf, M. and Wozniak, J.},
  title     = {{Workflows Community Summit: Bringing the Scientific Workflows 
               Research Community Together}},
  month     = {Mar},
  year      = {2021},
  publisher = {Zenodo},
  doi       = {10.5281/zenodo.4606958}
}
    \end{lstlisting}
    \\
    \vspace{0.5em}
    \textbf{License}
    \\
    This report is made available under a Creative Commons Attribution-ShareAlike 4.0 International license ({\scriptsize \url{https://creativecommons.org/licenses/by-sa/4.0/}}).
\end{tabular}
\end{table}
\vspace*{\fill}

\newpage

\includepdf[pages=-]{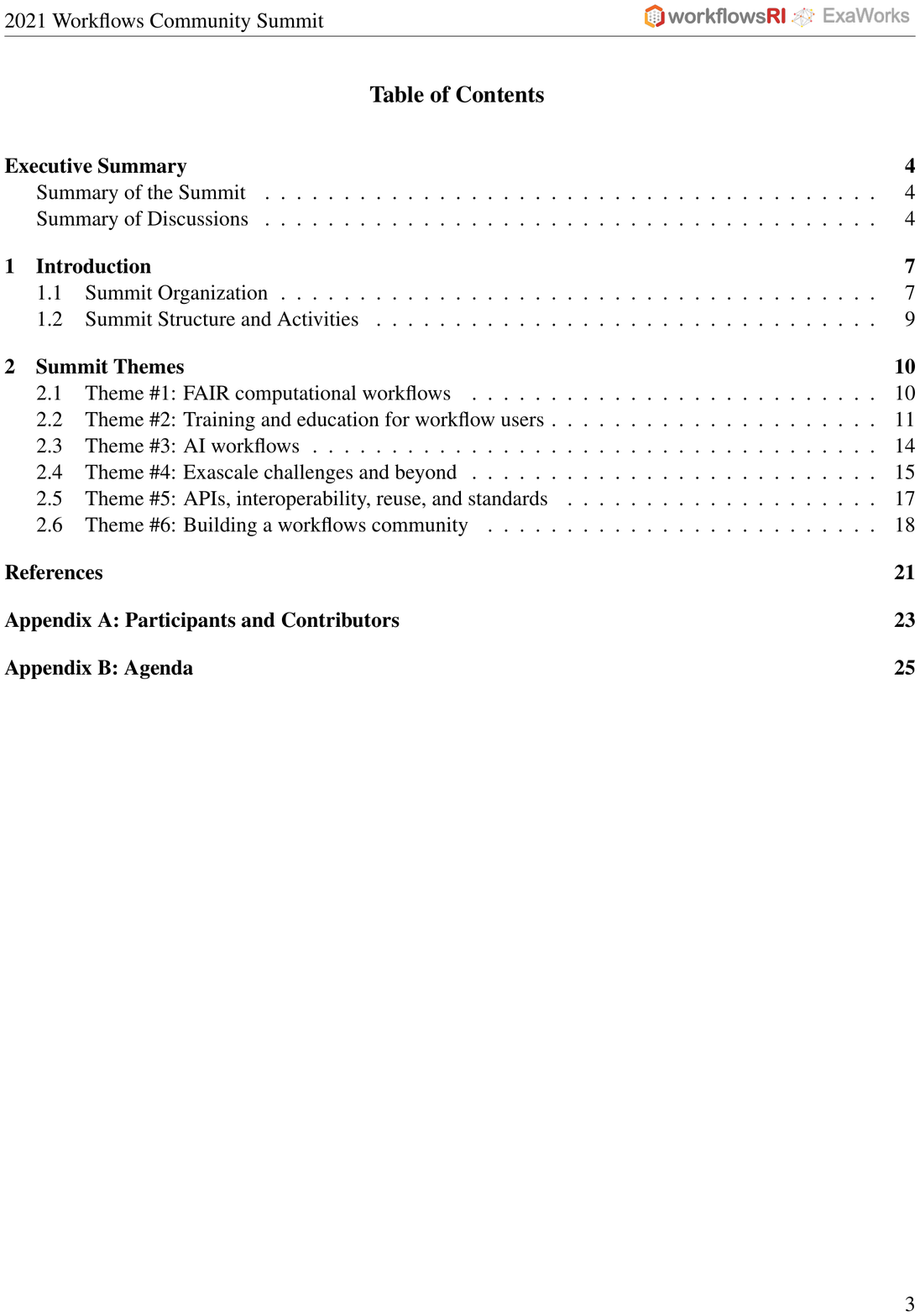}
\newpage

\cleardoublepage\phantomsection\addcontentsline{toc}{section}{Executive Summary}
\section*{Executive Summary}
\label{sec:summary}

Scientific workflows have been used almost universally across scientific domains, and have underpinned some of the most significant discoveries of the past several decades. Many of these workflows have high computational, storage, and/or communication demands, and thus must execute on a wide range of large-scale platforms, from large clouds to upcoming exascale high-performance computing (HPC) platforms. These executions must be managed using some software infrastructure. Due to the popularity of workflows,  workflow management systems (WMSs) have been developed to provide abstractions for creating and executing workflows conveniently, efficiently, and portably.  While these efforts are all worthwhile, there are now hundreds of independent WMSs, many of which are moribund. As a result, the WMS landscape is segmented and presents significant barriers to entry due to the hundreds of seemingly comparable, yet incompatible, systems that exist. As a result, many teams, small and large, still elect to build their own custom workflow solution rather than adopt, or build upon, existing WMSs. This current state of the WMS landscape negatively impacts workflow users, developers, and researchers.

\subsection*{Summary of the Summit}
\addcontentsline{toc}{subsection}{Summary of the Summit}

The ``Workflows Community Summit" was held online on January 13, 2021. The summit included 48 invited participants from a select group of international (Brazil, Germany, Spain, UK, and USA) lead researchers from distinct workflow management systems and users, and representatives from the National Science Foundation (NSF), and Department of Energy (DoE). The summit was co-organized by the PIs of two distinct projects: the NSF-funded WorkflowsRI project ({\small \url{https://workflowsri.org}}) and the DoE-funded ExaWorks project ({\small \url{https://exaworks.org}}).

The overarching goal of the summit was to develop a view of the state of the art and identify crucial research challenges in the workflow community. Prior to the summit, a survey sent to stakeholders in the workflow community (including both developers of WMSs and users of workflows) helped to identify key challenges in this community that were translated into 6 broad themes for the summit, each of them being the object of a focused discussion led by a volunteer member of the community. For each theme, the lead gave a plenary 5-minute lightning talk followed by focused discussions in breakout sessions. The goal of these sessions was to identify other challenges, if need be, and outline short- and long-term community efforts that could help tackle these challenges (summarized below).

This report documents and organizes the wealth of information provided by the participants before, during, and after the summit. Additional details (including the agenda and presentation videos) can be found at {\small \url{https://workflowsri.org/summits/community}}.

\subsection*{Summary of Discussions}
\addcontentsline{toc}{subsection}{Summary of Discussions}

\pp{Theme \#1: FAIR computational workflows}
The FAIR principles have laid a foundation for sharing and publishing digital assets and in particular data. It is thus important that scientific workflows both support the creation of FAIR data and themselves adhere to the FAIR principles. The group identified a number of core challenges: (i)~defining FAIR principles for computational workflows that consider the complex lifecycle from specification to execution and data products; (ii)~capturing and moving workflow components, dependencies, and application environments to enable reuse; (iii)~defining metrics to measure the FAIRness of a workflow, and labelling; and (iv)~engaging the community to define principles, policies, and best practices. Based on these challenges, the following short-term community efforts were identified: (i)~review prior and current efforts for FAIR data and software with respect to workflows; (ii)~outline rules for FAIR workflows; and (iii)~define recommendations for FAIR workflow repositories. As long-term community efforts, the group identified the following: (i)~define recommendations for workflow developers and workflow management systems in a view to supporting best practices for FAIR data; (ii)~ensure provenance can capture necessary information; and (iii)~automate FAIRness in workflows.

\pp{Theme \#2: Training and education for workflow users}
There is a strong need for more, better and/or new training and education opportunities for workflow users. In particular, many users ``re-invent the wheel'' without reusing software infrastructures and workflow tools that would make their workflow execution more convenient, more efficient, more evolvable, and more portable. The group identified a number of challenges: (i)~many workflow tools have high barrier to entry, sometimes make it almost needed to ``ship a developer along with the workflow tool''; (ii)~lack of training material and infrastructure (there are few ``recipes'' or ``cookbooks'' for workflow systems); (iii)~lack of standard and guidance for selecting appropriate workflow systems; (iv)~homegrown workflow solutions and constraints can prevent users from exactly reproducing the functionality of their homegrown system on workflow tools developed by others that would afford many other advantages; (v)~difficult to pick the appropriate time for providing training; and (vi)~unawareness of the workflow technological and conceptual landscape. Given these challenges, the group proposed the following short-term community efforts: (i)~identifying and advertising a set of basic sample workflow patterns; and (ii)~committing to training efforts in person. Long-term community efforts include: (i)~developing a community workflow knowledge-base that users can look up; (ii)~including workflow terminology and concepts in university curricula; (iii)~developing and contributing workflow modules to existing software carpentry efforts; and (iv)~looking at current research on technology adoption.

\pp{Theme \#3: AI workflows}
Scientific workflows empowered with machine learning (ML) techniques largely differ from traditional scientific workflows running on HPC machines. While scientific workflows  traditionally take few input data and produce large outputs, ML approaches read a considerable number of input files and usually produce one output, the trained model. One of the major differences is the inherent iterative nature of ML processes -- AI workflows often feature feedback loops. The group identified the following set of core challenges: (i)~lack of support for heterogeneity of compute resources (CPUs, GPUs, TPUs, etc.); (ii)~lack of support for fine-grained data management features (at the file level) and versioning features as well as adequate data provenance capabilities; (iii)~lack of capabilities for enabling workflow steering and dynamic workflow execution; and (iv)~integration of ML frameworks into the current HPC landscape. Based on these challenges, the following short-term community efforts were identified by the group: (i)~develop comprehensive use cases for sample problems with representative workflow structures and data types; and (ii)~define a systematic process for identifying and categorizing the challenges for enabling AI workflows. As a long-term community effort, it is suggested that AI workflows be developed as a way to benchmark HPC systems.

\pp{Theme \#4: Exascale challenges and beyond}
Given the computational demands of many workflows, it is crucial that their execution be not only feasible but also straightforward and efficient on large-scale HPC systems, and in particular upcoming exascale systems. The group identified many technical and non-technical challenges: (i)~resource allocation policies and schedulers are not designed for workflow-aware abstractions; (ii)~harmonization of workflow executions with non-workflow jobs in which workflow users tend to ``fake it'' by using an ill-fitted job abstraction; (iii)~unfavorable design of resource descriptions and mechanisms for workflow users/systems; (iv)~workflow-specific solutions for tackling fault-tolerance and fault-recovery are typically not readily available or deployed; and (v)~degree of complexity for advancing to exascale architectures (requirements/capabilities, testing, etc.). The following short-term community goals were identified by the group: (i)~develop documentation in the form of workflow templates/recipes/miniapps for execution on high- end HPC systems; (ii)~specify community benchmark workflows for exascale execution; and (iii)~include workflow benchmarks in exascale machines acceptability tests. As a possible longer-term community effort, the group suggested that workflow requirements be included very early on in the machine procurement process at compute centers.

\pp{Theme \#5: APIs, interoperability, reuse, and standards}
Each workflow system serves a different user community or underlying compute engine, albeit with substantial technical and conceptual overlap. Divergences include different workflow structures, different optimization goals, and different capabilities of execution systems. The group has identified the following challenges: (i)~there are broad differences between workflow systems that make interoperability hard to achieve; (ii)~WMSs have many different layers that may be able to interoperate, thus interoperability at some layers is likely to be more impactful than others; (iii)~most efforts to standardize WMSs and components have led to the definition of a ``standard'' developed by a subset of the community; and (iv)~it is difficult to quantify the value in having common representations of workflows/tasks. Short-term community efforts envisioned by the group include: (i)~compare differences and commonalities between different systems; (ii)~identify and characterize domain-specific efforts; (iii)~develop case-studies of business process workflows and serverless workflow systems; and (iv)~identify workflow patterns. The following long-term community effort was identified: sustained funding for workflow standards.

\pp{Theme \#6: Building a workflows community}
Given the current size and fragmentation of the workflow technology landscape, there is a clear need to establish a cohesive community of workflow developers and users. Specific challenges identified by the group include: (i)~defining what is meant by a ``workflow community''; (ii)~remedying the current inability to link developers and users to bridge translational gaps; (iii)~creating a formal network of researchers, developers, and users; and (iv)~providing pathways for participation. Based on these challenges, a short-term community effort identified by the group would be to establish a common knowledge-base for workflow technology. As a long-term community effort the group proposed the establishment of a ``Workflow Guild'', i.e., an organization focused on interaction and good relationships and self-support between subscribing workflow developers and their systems.

\newpage

\section{Introduction}
\label{sec:introduction}

Scientific workflows have been used almost universally across scientific domains, and have underpinned some of the most significant discoveries of the past several decades. Many of these workflows have high computational, storage, and/or communication demands, and thus must execute on a wide range of large-scale platforms, from large clouds to upcoming exascale HPC platforms~\cite{ferreiradasilva-fgcs-2017}. These executions must be managed using some software infrastructure. Historically, such approaches have tended to consist in complex, integrated  software systems, developed in-house by workflow practitioners using a range of legacy technologies (even including sets of ad-hoc scripts!).  Due to the popularity of workflows,  workflow management systems (WMSs) have been developed to provide abstractions for creating and executing workflows conveniently, efficiently, and portably.  While these efforts are all worthwhile, there are now hundreds of independent WMSs~\cite{workflow-systems}, thousands of researchers and developers, and a rapidly growing corpus of workflows research publications. The WMS technology landscape is thus segmented and presents significant barriers to entry due to the hundreds of seemingly comparable, yet incompatible, systems that exist. Another fundamental problem is that there are conflicting theoretical bases and abstractions for a WMS. Systems that use the same underlying abstractions can likely be translated between, which is not the case for systems that use different abstractions. More specifically, each system has a layered model that abstractly underlies it: (i)~if these models are the same for two systems, they are compatible to some extent, and if they implement the same layers, they can be interchanged (it may require some translation effort); (ii)~if these models are the same for two systems, but they implement different layers, they can be complementary, and may have common elements that could be shared; (iii)~if the models are distinct, the WMSs might not be exchangeable or interoperable. As a result, many teams, small and large, still elect to build their own custom workflow solution rather than adopt, or build upon, existing WMSs. This current state of the WMS landscape negatively impacts workflow users, developers, and researchers~\cite{deelman2018future}.

\subsection{Summit Organization}

\begin{figure}[!t]
    \centering
    \includegraphics[width=\linewidth]{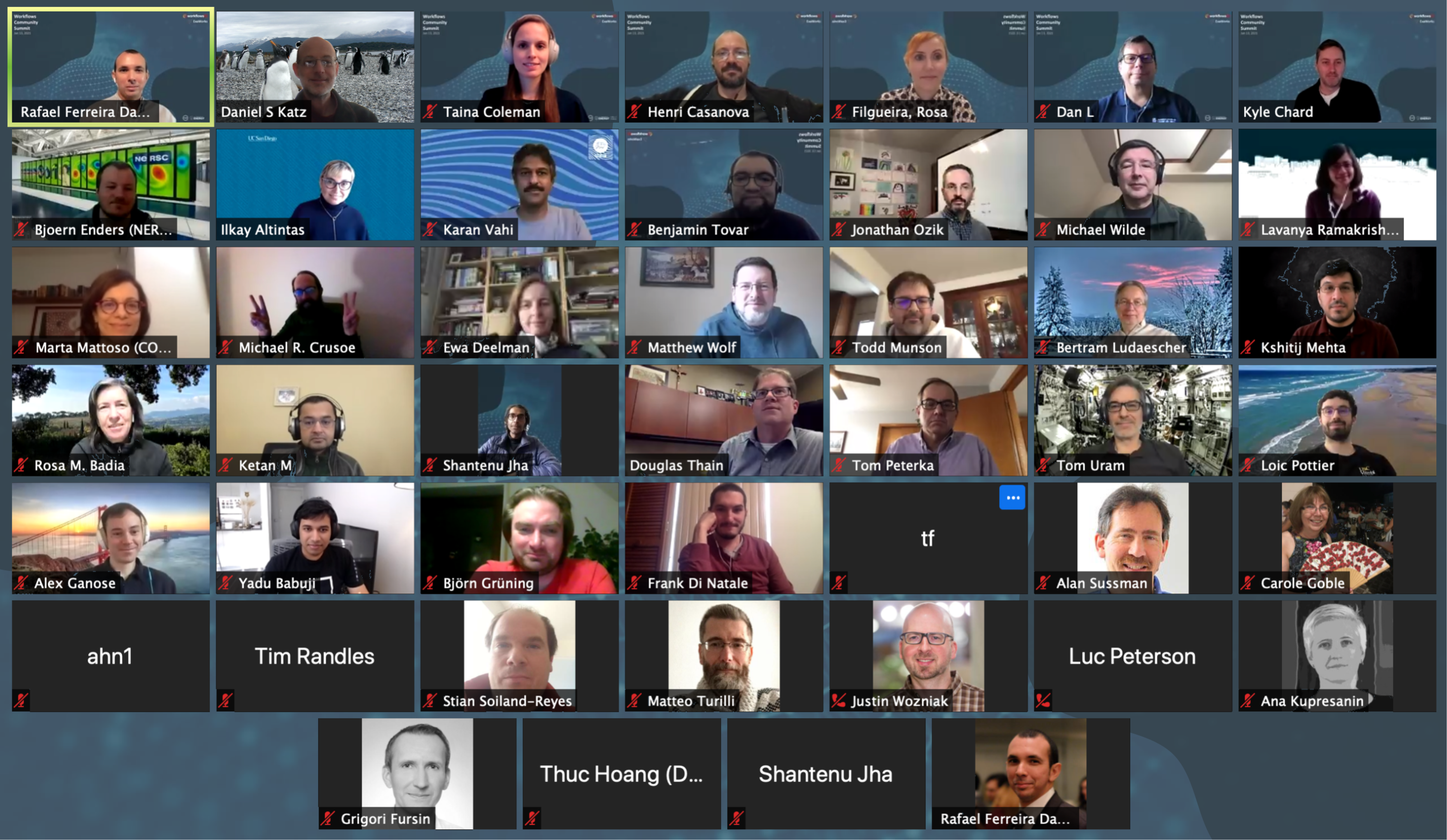}
    \caption{Screenshot of the Workflows Community Summit participants. (The event was held virtually via Zoom.)}
    \label{fig:participants}
\end{figure}

This document reports on discussions and findings from an international ``Workflows Community Summit" that took place on January 13, 2021~\cite{wcs}. 
The summit included 48 invited participants from a select group of international (Brazil, Germany, Spain, UK, and USA) lead researchers from distinct workflow management systems and users, and representatives from the National Science Foundation (NSF), and the Department of Energy (DoE) (Figure~\ref{fig:participants}).

The summit was co-organized by the PIs of two distinct projects: the NSF-funded WorkflowsRI project~\cite{workflowsri} and the DoE-funded ExaWorks project~\cite{exaworks} (Table~\ref{tab:organization}). These two projects aim to develop solutions to address some of the challenges faced by the workflow community, albeit in different ways. WorkflowsRI focuses on defining the blueprints for a community research and development infrastructure. Specifically, the goal is to bring together workflow users, developers, and researchers to identify common ways to develop, use, and evaluate workflow systems, resulting in a useful, maintainable, and community-vetted set of principles, methods, and metrics for WMS research and development. The resulting research infrastructure will provide an academic nexus for addressing pressing workflow challenges, including those resulting from the increasing heterogeneity and size of computing systems, changing workflow patterns to encompass rapid feedback and machine learning models, and the increasing diversity of the user community. ExaWorks aims to develop a multi-level workflows SDK that will enable teams to produce scalable and portable workflows for a wide range of exascale applications. ExaWorks does not aim to replace the many workflow solutions already deployed and used by scientists, but rather to provide a robust SDK and work with the community to identify well-defined and scalable components and interfaces that can be leveraged by new and existing workflows. A key goal is that SDK components should be designed to be usable by many other WMS, thus facilitating software convergence in the workflows community. Although clearly distinct, these two projects have similar motivations, and their planned research and development activities could benefit from collaboration and even co-design.

\begin{table}[!t]
    \centering
    \setlength{\tabcolsep}{12pt}
    \begin{tabular}{lll}
        \toprule
        \textbf{Project} & \textbf{Name} & \textbf{Affiliation} \\
        \midrule
        WorkflowsRI & Rafael Ferreira da Silva & University of Southern California \\
        & Kyle Chard & University of Chicago \\
        & Henri Casanova & University of Hawai'i at M\~anoa \\
        & Tain\~a Coleman & University of Southern California \\
        \rowcolor[HTML]{F2F2F2}
        ExaWorks & Dan Laney & Lawrence Livermore National Laboratory \\
        \rowcolor[HTML]{F2F2F2}
        & Dong H. Ahn & Lawrence Livermore National Laboratory \\
        \rowcolor[HTML]{F2F2F2}
        & Kyle Chard & University of Chicago \\
        \rowcolor[HTML]{F2F2F2}
        & Shantenu Jha & Rutgers University \\
        \bottomrule
    \end{tabular}
    \caption{Workflows Community Summit organizers.}
    \label{tab:organization}
\end{table}

\subsection{Summit Structure and Activities}

\begin{wrapfigure}{r}{0.5\textwidth}
    \centering
    \includegraphics[width=0.48\textwidth]{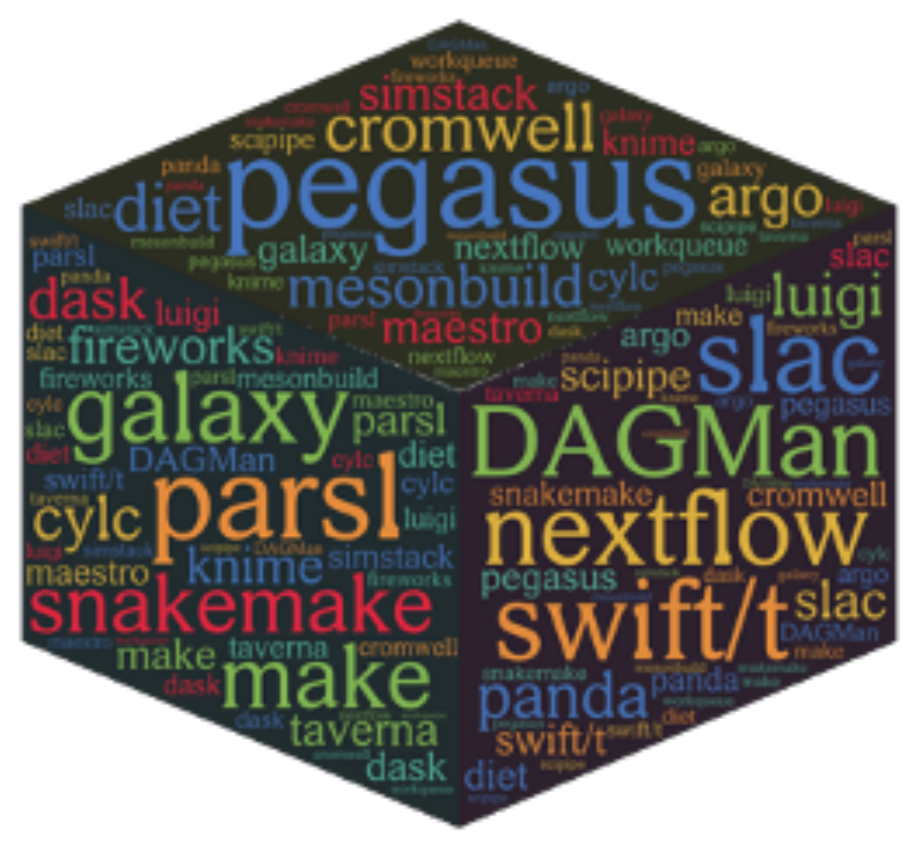}
    \caption{Word box of used WMSs obtained from the Community Research Infrastructure Survey.}
    \label{fig:word-box}
\end{wrapfigure}

Prior to the summit, we conducted a survey (``Community Research Infrastructure Survey") sent to stakeholders in the workflow community, including both developers of WMSs and users of workflows (and possibly WMSs). The goal of this survey was to to better understand workflow research and development problem domains, requirements for WMSs, capabilities and perceived limitations of current WMSs, and opportunities for supporting and improving workflow and WMS research and development activities. To date we have gathered responses from 73 participants  from a diverse set of WMSs developers and users (40 self-identified as developers and 33 as users). We have (and will continually) reached out to a number of lead researchers from the workflows community to assist in disseminating the survey with their collaborators. Figure~\ref{fig:word-box} shows a word box underlining the different sets of WMSs used by the survey participants. Although the survey is still active, we believe the responses already gathered capture the major state of the art of workflow research and challenges.

The survey asked a wide range of questions regarding WMS design and use. But one of its goals was also to get a sense of the key challenges on which the workflows community should focus its efforts.  Survey participants were asked to respond with free-text descriptions of the most pressing challenges from their own perspective. Based on survey answers to the challenge question, we identified six broad themes for the summit, each of which was the object of a focused discussion led by a volunteer community member (Table~\ref{tab:lead}).

\begin{table}[!t]
    \centering
    \begin{tabular}{l|l}
        \toprule
        \textbf{Theme} & \textbf{Discussion Lead} \\
        \midrule
        Theme \#1: FAIR computational workflows & Carole Goble, Univ. of Manchester, UK \\
        \rowcolor{Gray}
        Theme \#2: Training and education for workflow users & Lavanya Ramakrishnan, LBNL, USA \\
        Theme \#3: AI workflows & Luc Peterson, LLNL, USA \\
        \rowcolor{Gray}
        Theme \#4: Exascale challenges and beyond & Bjoern Enders, NERSC, USA \\
        Theme \#5: APIs, interoperability, reuse, and standards & Douglas Thain, Univ. of Notre Dame, USA \\
        \rowcolor{Gray}
        Theme \#6: Building a workflows community & Ilkay Altintas, SDSC, USA \\
        \bottomrule
    \end{tabular}
    \caption{Workflows Community Summit themes and discussion leaders.}
    \label{tab:lead}
\end{table}

For each of the themes shown in Table~\ref{tab:lead}, the lead gave a plenary 5-minute lightning talk, highlighting the theme and its challenges (videos for each presentation can be found at the WorkflowsRI's YouTube channel~\cite{workflowsri-youtube}). Each theme was then discussed in breakout sessions, with the lead coordinating the discussion and a shared document being used to capture notes and for asynchronous discussion. The goal of these sessions was to identify other challenges, if need be, and outline short- and long-term community efforts that could help tackle these challenges.  The lead then reported on the outcome of the discussion in plenary sessions, after which final remarks were given by the organizers and the summit was adjourned.  In the following sections, we summarize the outcome from the breakout session for each theme.

\section{Summit Themes}

\subsection{Theme \#1: FAIR computational workflows}
\label{sec:theme1}

The FAIR principles~\cite{wilkinson2016fair} have laid a foundation for sharing and publishing digital assets and in particular data. The FAIR principles emphasise machine accessibility and that all digital assets should be Findable, Accessible, Interoperable, and Reusable. Scientific workflows encode the methods by which the scientific process is conducted and via which data are created. It is thus important that scientific workflows both support the creation of FAIR data and themselves adhere to the FAIR principles. The group identified a set of core challenges in this space. 

\begin{itemize}
    \item \textbf{What are the FAIR principles for Computational Workflow Objects?}
    Workflows are a unique type of digital asset that can be considered as data or software, or some combination of both. As such, there are a range of considerations with respect to the FAIR principles~\cite{goble2020}. Some perspectives are already well explored in data/software FAIRness, such as processal metadata, software metrics, and versioning; however, workflows create unique challenges such as representing a complex lifecycle from specification, to execution via a workflow management system, through to the data created at the completion of the workflow.


    \item \textbf{How do we design for and support reuse?}
    One of the most challenging aspects of making workflows FAIR is ensuring that they can be reused. These challenges include being able to capture and then move workflow components, dependencies, and application environments in such a way as to not affect the resulting execution of the workflow. The group suggested that further work is first required to understand use cases for reuse, before exploring methods for capturing necessary context and enabling reuse in the same or different environments. 

    \item \textbf{How do we measure and label a workflow as FAIR?}
    There are many metrics and features that could be considered in the decision of whether a workflow is FAIR. These features may well be different depending on the type of workflow and its application domain. Prior work in data and software FAIRness provides a starting point, however, these metrics will need to be updated for workflows. In terms of labeling, there has been widespread adoption of reproducibility badges for publications and of FAIR labels for data in repositories. Similar approaches could be applied to computational workflows. 

    \item \textbf{How do we work together?} 
    Developing methods for FAIR workflows requires community engagement to define principles, policies, and best practices; to standardize metadata representation and collection processes; to create developer-friendly guidelines and workflow-friendly tools; and to develop shared infrastructure for enabling development, execution, and sharing of FAIR workflows (e.g., via registries like WorkflowHub.eu~\cite{workflowhub-eu}).

\end{itemize}

\medskip
\noindent
Based on these challenges, the following \textbf{short-term community efforts} were identified by participants. 

\begin{enumerate}
    \item \textbf{Review prior and current efforts for FAIR data and software with respect to workflows.}
    Given the amount of prior work in this space, it is important to first understand what efforts could be adapted to workflows problems. Specific activities might include participating in the FORCE11/RDA/ReSA FAIR for Research Software (FAIR4RS) group~\cite{fair4rs} from the point of view of workflows as software and starting a task force (potentially tied to FAIR notebooks).
    
    \item \textbf{Outline rules for FAIR workflows.}
    Collect a set of real use cases and workflows in several domains to examine from the perspective of the FAIR data principles. This exercise will likely highlight areas in which the FAIR data principles adequately represent challenges in workflows. Based on these experiences, a set of simple rules could be defined for creating FAIR workflows (e.g., Nine Best Practices for Research Software Registries and Repositories: A Concise Guide~\cite{monteil2020nine}).
    
    \item \textbf{Define recommendations for FAIR workflow repositories.} 
    Building from the rules established for creating FAIR workflows and working
    with prominent workflows repositories (e.g., the WorkflowHub.eu registry~\cite{workflowhub}), 
    communities (e.g., the WorkflowHub Club), and workflow management systems
    define a set of recommendations that can be enacted by workflow repositories
    to support the development and sharing of FAIR workflows. 
    
\end{enumerate}

\medskip
\noindent
The following \textbf{long-term community efforts} were also proposed. 

\begin{enumerate}

    \item \textbf{Define recommendations for workflow developers and workflow management systems in a view to supporting best practices for FAIR data.} 
    Workflows play a crucial role in the creation of data and thus provide an opportunity to improve the FAIRness of data. These efforts relate not only to the workflow itself, but the workflow components, execution environments, and the different types of data. 
    
    
    \item \textbf{Ensure provenance can capture necessary information.} 
    There are many existing provenance and tracing standards developed for various purposes. We may need to implement or extend these standards to ensure they capture the information needed for FAIR workflows and to ensure that they can be used \emph{everywhere}.
    
    \item \textbf{Automate FAIRness in workflows.} 
    FAIR principles are more likely to be followed if the capture of metrics are automated and embedded in workflow management systems and languages (e.g., CWL annotations~\cite{cwl-annotations}). In which case, a workflow execution will become FAIR by default, or perhaps with minimal user curation.

\end{enumerate}

\subsection{Theme \#2: Training and education for workflow users}
\label{sec:theme2}

There is a strong need for more, better and/or new training and education opportunities for workflow users. In particular, many users ``re-invent the wheel" without reusing software infrastructures and workflow tools that would make their workflow execution more convenient, more efficient, more evolvable, and more portable. Specific challenges identified by the lead during their plenary presentation and the group during the breakout session include:


\begin{itemize}

    \item {\bf Many workflow tools have high barrier to entry.} 
    Workflow tools require a large amount of effort and time by the users due to a steep learning curve.  A contributing factor is that users may not know the required terminology and concepts. As a result, some have noted that what would be needed in the current technology landscape is to ``ship a developer along with the workflow tool", which is clearly not (always) feasible and definitely not scalable.

    \item{\bf Training material and infrastructure are lacking.} 
    One of the reasons for the above challenge is that there are few ``recipes" or ``cookbooks" for workflow systems. Ideally, when joining an organization and starting to use resources at some facility, there should be a way for users to be pointed to (online) resources with simple examples of how to use particular workflow systems. Furthermore, given that workflows and their execution platforms are complex and diverse, in addition to mere training material, there is a need for a training infrastructure that consists of workflows and accompanying data (small enough to be used for training purposes but large enough to be meaningful) as well as execution testbeds for running these workflows.

    \item {\bf It is difficult to choose and commit to a workflow tool.}
    Given the enormous number of workflow tools~\cite{workflow-systems}, and the lack of any standards, it is not easy for users to pick appropriate systems for their needs. But, perhaps more importantly, there is an understandable fear of being locked into a tool that at some point in the near future is no longer supported. Although documentation can be a problem, often it is guidance that is the more crucial issue. 

    \item {\bf Many users have their own imperfect, but working, homegrown workflow solutions.} 
    Many users have the necessary basic skills to execute a simple workflow on some system (e.g., using bash scripts and Slurm). Initial user attitude is thus often ``why should I learn anything new?" (also due to the above challenge). As requirements gradually increase, many users evolve their simple approach in ad-hoc ways, thus developing/maintaining a working but imperfect homegrown system.  There is thus a high risk of hitting technological or labor-intensiveness roadblocks, which could be remedied by using a workflow tool developed by others. But when ``graduating" to such a system, there will likely be constraints that prevent users from exactly reproducing the functionality of their homegrown system. The benefits of using the tool should thus largely outweigh the drawback of these constraints. Making a good case that it will is difficult because every user and use case is different.

    \item {\bf Providing training to users at the right time is not easy.}
    Given all the above challenges, it is not easy to reach out to users (providing there are mechanisms for doing so) at the appropriate time. Reach out too early and users will not view using a particular tool as compelling.  Reach out too late, and users are already locked into their homegrown system, even though in the long run this system will severely harm their productivity.

    \item {\bf Users may not be aware of the workflow technological and conceptual landscape.} Many users do not know that workflow tools even exist (which is a reason why users start developing their own ad-hoc approaches). More generally, users may not even know that workflows are a common model of computation for which knowledge, technology, and a community exist.

\end{itemize}

\medskip
\noindent
Although addressing all above challenges is an enormous undertaking, the following {\bf short-term community efforts} have been outlined by summit participants:

\begin{enumerate}
    
    \item \textbf{Identify and advertise a set of basic sample workflow patterns.}
    Lowering the entry barrier is key for enabling the next-generation of researchers to benefit from workflow systems. An initial approach would be to provide a basic set of simple, yet conceptually rich, (e.g., ``hello world" one-task workflows, chain workflows, fork-join workflows, simple dynamic workflows, all with a few ways of handling data and I/O, and all with a few target execution platforms). Then workflow systems can develop a documentation page, using a common format/framework, that shows how to run these patterns with their system. Summit participants noted that:
    \begin{compactitem}
        \item NERSC already hosts workflow tool pages for their users~\cite{nersc-workflows}, and could serve as inspiration for a larger, community effort;
        \item The above proposed material, or links to it, could be hosted on a community Web site;
        \item This material should be a mix of documentation and video tutorials.
    \end{compactitem}

    \item \textbf{Committing to training efforts in person.} 
    Mechanisms should be identified at institutional level to commit workflow systems training efforts in person, in addition to the above training material. Summit participants noted that:
    \begin{compactitem}
        \item This should be based on existing facilities and universities efforts that already exist, which need to be identified;
        \item The scope of the training should be narrowed down so it's manageable;
        \item The issue of ``who trains the trainers?" needs to be addressed.
    \end{compactitem}

\end{enumerate}

\noindent
Summit participants also outlined the following possible {\bf longer-term community efforts}:

\begin{enumerate}
  \item \textbf{Develop a community workflow knowledge-base that users can look up.} 
  There is a established community of workflow researchers, developers, and users that have extensive expertise knowledge regarding specific tools, systems, applications, among others. It is crucial to capture such knowledge and bootstrap a community workflow knowledge-base (following standards for documentation, interoperability, etc.) for training and education, so that we would not have to ``ship the developer with every tool".
  
  \item \textbf{Include workflow terminology and concepts in university curricula.}
  In an attempt to address the issue of reaching out to researches/users at the right time, workflow concepts should be taught at early stages of the researchers/users education path. Specifically, introducing such concepts as part of the university curricula, including domain science curricula, could be achieved by leveraging modeling and simulation (e.g., eduWRENCH pedagogic modules~\cite{tanaka2019eduhpc, eduwrench}).
  
  \item \textbf{Develop and contribute workflow modules to existing software carpentry efforts.}
  Software Carpentry~\cite{swcarpentry} is a volunteer organization whose goal is to make scientists more productive and their work more reliable by teaching them basic computing skills. Developing modules that provide basic concepts of workflows and distributed computing would significantly lower the entry barrier for adopting WMSs as part of the research process.
  
  \item \textbf{Look at current research on technology adoption.} 
  The workflow research community would benefit from collaborations with social scientists and sociologists so as to help define an overall strategy for approaching some of the above challenges.
\end{enumerate}

\subsection{Theme \#3: AI workflows}
\label{sec:theme3}

Artificial intelligence (AI) and machine learning (ML) techniques are becoming increasingly popular within the scientific community. ML techniques are very efficient to analyze the vast amount of data generated by large-scale simulations. Scientific workflows empowered with ML techniques largely differ from traditional scientific workflows running on high-performance computing (HPC) machines. While scientific workflows (i.e., one large-scale application simulating a scientific object or process) traditionally take few input data and produce large outputs, ML approaches read a considerable number of input files and usually produce one output, the trained model. Additionally, scientific workflows traditionally exhibit regular data access patterns whereas ML-empowered workflows show more randomness in access pattern, caused by the sampling of input files mandatory for achieving convergence of the training process. Finally, one of the major differences is the inherent iterative nature of ML processes -- AI workflows often feature feedback loops. These differences pose several challenges to the workflow community, and specific challenges identified by the summit participants include:

\begin{itemize}
    \item \textbf{Support for heterogeneity of compute resources.} 
    AI techniques run more efficiently on massive parallel architectures such as GPUs and tensor processing units (TPU). WMSs aiming at better support workflows with AI sub-components need to properly manage heterogeneous resources, offload heavy computations to GPUs, and manage data between GPU and CPU memories.
    
    \item \textbf{Fine-grained data management and versioning.} 
    AI workflows manipulate a large number of files which are sampled multiple times during the training phase. Scientists require capabilities to track files movements, assess which part of the data set has been used and when. This will help understand training performance and accuracy. Data is the cornerstone of modern AI approaches. For example, many ML practitioners often wish to run multiple times the same workflow on different data sets, or to re-run the workflow with an augmented data set. To face these challenges, WMSs should provide fine-grained data management features (at the file level) and versioning features as well as adequate data provenance capabilities.
    
    \item \textbf{Dynamic workflow steering.} 
    Most AI/ML approaches are iterative, i.e. an AI workflow iterates over a data set to train a model, at each pass on the data set the training accuracy is verified. With such an iterative approach, users typically want to interact with the execution of the workflow at runtime, e.g., start a new ML workflow with different input parameters if the accuracy of the results computed by the previous workflow execution is not above some threshold. 
    
    \item \textbf{Dynamic workflow execution.} 
    By design, ML-empowered workflows are dynamic, in contrast to traditional workflows with more structured and deterministic computations. The execution of ML tasks typically depends on a large number of parameters, called hyperparameters in the ML literature. Thus, it is common practice to explore the hyperparameters space leveraging frameworks for hyperparameter optimization (HPO)~\cite{akiba2019optuna, bergstra2013hyperopt, wozniak2018}. At runtime, the workflow execution graph can potentially evolve based on internal metrics (e.g., accuracy), which may increase the number of tasks in the workflow graph or trigger task preemption. In some sense, the shape of the workflow (i.e., the DAG) is partially defined by these hyper-parameters. Workflow systems should thus support dynamic branching (e.g., conditionals, criteria) and partial workflow re-execution on-demand.
    
    \item \textbf{ML frameworks integration into the current HPC landscape.} 
    Modern AI and ML development is not driven by the HPC community. Most of the widely-used ML frameworks come from industry and have been designed to run on cloud infrastructures. ML frameworks use Python and R-based libraries and do not follow the classic HPC model: C/C++/Fortran, MPI, and OpenMP, and submission to an HPC batch scheduler. Yet, some efforts in the HPC community such as LBANN~\cite{lbann} (Lawrence Livermore National Laboratory (LLNL)) for HPC-enabled ML, and EMEWS~\cite{ozik2016} for integrating ML algorithms to control HPC workflows (Argonne National Laboratory), have already started to address some of the above issues. However, there is a clear disconnect between HPC motivations, needs, and requirements, and AI/ML current practices.
    
\end{itemize}

\medskip
\noindent
Based on these challenges, the following \textbf{short-term community efforts} were identified by participants:

\begin{enumerate}
    \item \textbf{Develop comprehensive use cases.}
    To address the disconnect between HPC systems and practices and the AI workflows, the community needs to develop sets of example use cases for sample problems with representative  workflow structures and data types. In addition to expanding upon the above challenges, the community could ``codify'' these challenges in these example use cases.

    \item \textbf{Systematically expand on challenges.}
    The set of challenges for enabling AI workflows is extensive. The community needs to define a systematic process for identifying and categorizing these challenges. A short-term recommendation would be to write a community white paper about AI Workflow challenges/needs.
\end{enumerate}

\medskip
\noindent
The following \textbf{long-term community effort} was also proposed:

\begin{enumerate}
    \item \textbf{Produce AI workflows for benchmarking HPC systems.} 
    Building from the use cases established above for the needs and requirements of AI workflows define AI-Workflow mini-apps, which could be used to pair with vendors/future HPC developers so that the systems can be benchmarked against these workflows, and therefore support the co-design of emerging or future systems (e.g., ML Commons~\cite{mlcommons} and Collective Knowledge (CK)~\cite{ck2021}).
\end{enumerate}

\subsection{Theme \#4: Exascale challenges and beyond}
\label{sec:theme4}

Given the computational demands of many workflows, it is crucial that their execution be not only feasible but also straightforward and efficient on large-scale HPC systems, and in particular upcoming exascale systems. Many technical and non-technical challenges were identified by the lead during the plenary presentation and the group during the breakout session, including:



\begin{itemize}
    \item {\bf Interaction with resource allocation policies and schedulers.}
    HPC resource allocation policies and schedulers are typically not designed with workflow applications in mind, and typically provide a mere ``job" abstraction instead of workflow-aware abstractions. Workflow users/systems are thus left to their own devices to make their workflows run on top of this ill-fitted abstraction. This abstraction also makes it difficult to control low-level behavior that is critical to workflows (i.e., precise mapping of tasks to specific compute resources on a compute node. Furthermore, there is a clear lack of support for elasticity (i.e., scaling up/down the number of nodes). Overall, it is currently difficult to run workflows efficiently and conveniently on HPC systems without extending (or overhauling?) resource management/scheduling approaches, which ideally would allow programmable, fine-grain application-level resource allocation and scheduling.

    \item {\bf Harmonization of workflow executions with non-workflow jobs.}
    Related to the above challenge, it is currently not possible to support both workflow and non-workflow users well on the same system. The typical approach is for workflow users to ``fake it" by using the ill-fitted job abstraction. Some features needed by workflows are often available. For instance, batch schedulers can support elastic jobs (e.g., Slurm). But experience shows that system administrators may not be keen on enabling this, as they deem long static allocations preferable. A cultural change is perhaps needed as it seems that workflows are not yet considered as high-priority applications by high-end compute facilities administrators (or perhaps many of these administrators are simply not really aware of workflows).

    \item {\bf Benefiting from accelerators and the like.} 
    Hybrid architectures are key to high performance and many workflows can or are specifically designed to exploit them. However, on HPC systems, the necessary resource descriptions and mechanisms are not necessarily available to workflow users/systems (even though some workflow systems have successfully interfaced to such mechanisms on particular systems). Although typically available as part of the ``job" abstraction, it is often not clear how a workflow system can discover and user them effectively.

    \item {\bf Workflow-specific fault-tolerance issues.} 
    Fault-tolerance and fault-recovery are well-known issues on exascale system, with a large amount of work and working solutions for traditional parallel jobs. Fault-tolerance and fault-recovery techniques for workflows have been the object of several research works, but workflow-specific solutions are typically not readily available or deployed. This has to change if workflows are to run feasibly and efficiently on exascale platforms.

    

    


    \item {\bf Graduating to exascale architectures.} 
    Workflows are built on smaller platforms, but when moving up to exascale, how does one express new requirements/capabilities and deal with new constraints?  How does one test before deployment (what is a ``local" exascale workflow?)

\end{itemize}

\medskip
\noindent
Although addressing all above challenges is an enormous undertaking, the following three {\bf short-term community efforts} have been outlined by summit participants:

\begin{enumerate}
    \item Develop documentation in the form of \textbf{workflow templates/recipes/miniapps for execution on high-end HPC systems}, which would be hosted on a community web site. Summit participants noted that some efforts are already underway that provide partial solutions (e.g., nf-core~\cite{nfcore}). For instance, collections of workflows exist but do not come with any execution capabilities, say on community testbeds. Some compute facilities, such as NERSC, provide workflow tool documentation and help with their users~\cite{nersc-workflows}. These solutions should be cataloged as a starting point.  One idea also could be for HPC facilities to organize yearly ``workflow days", in which they give workflow users and developers early access to machines to try out their workflows, thus gathering feedback from users and developers.

    \item Specify \textbf{community benchmark workflows for exascale execution}, exerting  all relevant hardware as well as functionality capabilities. Then it becomes possible for different workflow systems to execute these benchmarks. Summit participants noted that several efforts already exist that could provide initial ideas (e.g., OpenEBench~\cite{openebench}).

    \item Include workflow benchmarks in {\bf exascale machines acceptability tests}. Summit participants notes that there will be a need to pick particular workflow systems to run these benchmarks, which will  require educating HPC folks about these tools.
\end{enumerate}

\medskip
\noindent
Summit participants also outlined the following possible {\bf longer-term community effort}:

\begin{enumerate}

    \item Include workflow requirements very early on in machine procurement process for machines at compute centers (which will require miniapps and/or benchmark specifications, as well as API/scheduler specifications,  as outlined in the above proposed short-term efforts).


\end{enumerate}

\subsection{Theme \#5: APIs, interoperability, reuse, and standards}
\label{sec:theme5}

There has been an explosion of workflow (orchestration) technologies in the last decade~\cite{workflow-systems}. Each workflow system serves a different user community or underlying compute engine, albeit with substantial technical and conceptual overlap.  There are of course many good reasons for divergence, including that use cases have completely different workflow structures, resources have very different optimization goals, and execution systems provide fundamentally different capabilities. However, this divergence also leads to missed opportunities for interoperability. For example, it is often difficult for workflows to be ported between systems, for provenance to be captured in similar ways, and for workflow systems to leverage different execution engines, schedulers, or monitoring services. 


\begin{itemize}
    \item \textbf{There are broad differences between workflow systems.}
    Workflow systems often grow organically, developers start by solving a problem and they end up with a new workflow tool. In some cases, workflow systems may differ by design, rather than by accident.  For example, they offer fundamentally different abstractions or models for a workflow: graph vs cycles vs imperative or control flow vs data flow vs control and data flow. These fundamental differences make it such that interoperability may not even be possible. For example, when workflow systems apply different abstractions, it may not be possible to interoperate/convert in  an automated way.

    \item \textbf{At which layer(s) should we focus on interoperability?}
    Workflow systems have many different layers that may be able to interoperate, for example, workflow description, task description,  task definition, task submission, data passing, etc. Interoperability at some layers is likely to be more impactful than others. Further, interoperability does not need to imply agreement and for workflow systems to implement a standard interface; instead, it may occur via shim layers or intermediate representations, in a similar manner to compiling to a high level language.

    \item \textbf{There are many non-technical challenges that limit interoperability.} 
    Most efforts to standardize workflow systems and components have led to the definition of a ``standard'' developed by a subset of the community. This often serves to create $n+1$ version of that component or system and requires that other systems conform to that specification. 

    \item \textbf{Is there value in having common representations of workflows/tasks?} 
    One of the most obvious areas for agreement is in representation of workflows and their tasks. However, workflow systems have very different models (e.g., dynamic vs static graphs or data flow vs control flow) which make it difficult to share representations and attempts to standardize may lead to overly generic interfaces which inhibit usability.
    
    \item \textbf{Where should we start?}
    One area of agreement is the need for a common submission model for heterogeneous platforms. The differences between schedulers and systems is a universal challenge faced by workflow developers. Further challenges relate to authentication and authorization models deployed on many systems (e.g., two factor authentication). A new JLESC project might be a model~\cite{JLESC-common-registry}.

\end{itemize}

\medskip
\noindent
Based on these challenges, the following \textbf{short-term community efforts} were identified by participants:

\begin{enumerate}
    \item \textbf{Compare differences and commonalities between different systems.} 
    Host several ``bake-off'' to compare systems, including for task and workflow definitions as well as job execution interface. Engage participants to write and execute these workflows and identify commonalities between systems. A successful example of this approach can be seen with the GA4GH-DREAM challenge~\cite{ga4gh}. 
    

    \item \textbf{Identify and characterize domain-specific efforts.}
    There is likely to be greater opportunities for standardization within domains, and indeed some domains have already made good progress here. By reviewing these areas we can determine what has worked and what has not worked while also providing the opportunity to build on successful prior efforts. 
    

    \item \textbf{Case study of business process workflows.}
    Identify a set of community suggested use cases for connecting a business process to a batch workflow. Implement those connections and write up the lessons learned.

    \item \textbf{Case study of serverless workflow systems.}
    Identify a set of community suggested use cases that are implemented with one of the emerging FaaS (function-as-a-service) workflow systems provided by industry (e.g. AWS Step Functions, Azure Durable Functions, Google Cloud Functions, IBM Composer) and compare them against an implementation with a popular scientific workflow system. Such a comparison may turn out complementary features that can be of benefit for both industry and scientific workflow systems.
    
    \item \textbf{Identify workflow patterns.}
    Prior work has established a set of common workflow patterns, however, there is some uncertainty of the scope of these patterns (e.g., for representing patterns in dynamic workflows). We would benefit from surveying published patterns (e.g., Workflow Patterns~\cite{workflowpatterns}) and identifying gaps seen by the community. 

\end{enumerate}

\medskip
\noindent
The following \textbf{long-term community effort} was identified. 

\begin{enumerate}
    \item \textbf{Sustained funding for workflow standards.} 
    The current funding and research recognition models actively work against standardization. Developing sustained funding models for building and evolving workflow standards, encouraging their adoption, supporting interoperability, testing, and providing user and developer training would reduce these challenges. 
\end{enumerate}

\subsection{Theme \#6: Building a workflows community}
\label{sec:theme6}

Given the current size and fragmentation of the workflow technology landscape, there is a clear need to establish a cohesive community of workflow developers and users. The community could avoid unnecessary duplication of efforts and allow for sharing, and thus growth, of knowledge. To this end, there are four main components that need to be addressed for building a community: (i)~identity building, (ii)~trust, (iii)~participation, and (iv)~rewards. Specific challenges identified by the lead during their plenary presentation and the group during the breakout session include:

\begin{itemize}
    \item \textbf{Defining what we mean by a ``workflow community".}
    The most natural idea is to think of two distinct communities: (i)~a Workflow Research and Development Community and (ii)~a Workflow User Community. The former gathers people  who share interest in workflow research and development, and corresponding sub-disciplines.  Subgroups of this community are based on common methodologies, technical domains (e.g., computing, provenance, design), scientific disciplines, as well as geographical and funding areas. The latter gathers anyone using workflows for optimization of their work processes. However, people in this community may not themselves feel that they belong to it: most domain science users think of themselves in their specific disciplines first (and they just happen to use workflows to get their work done, rather than belonging to a perhaps elusive ``workflow user community'').
    
    \item \textbf{Making it possible to link developers and users to bridge translational gaps.}
    The two aforementioned communities are not necessarily disjoint, but currently have little overlap. And yet it is crucial that they interact.  Such interaction seems to happen only on a case-per-case basis, rather than via organized community efforts.  One could, instead, envision a single community (e.g., team of users, or ``team-flow'') that gathers both workflow system developers and workflow-focused users, and even compute platform folks, with the common goal of spreading knowledge and adoption of workflows, thus working towards increased sharing and convergence/interoperation of technologies and approaches.
    
    \item \textbf{Creating a formal network of researchers, developers, and users.}
    Establishing trust and processes is key for bringing both communities together. There is no one-size-fits-all workflow system or solution for all domains, instead each domain presents their own specific needs and have different preferred ways to address problems. There is then a pressing need for maintaining documentation and dissemination that fits different usage options and needs.
    
    \item \textbf{Providing pathways for participation.}
    Given the above, there is a number of existing community efforts that could serve as inspiration. Among these, Workflowhub.eu~\cite{workflowhub}, and in particular the WorkflowHub Club activity, and Galaxy, were cited as good examples of successful community efforts. A suggestion was to gather experience from computing facilities where teams have successfully adopted and are successfully running workflow systems (e.g., NERSC, ORNL.)  Another possibility is to use proposal/project reviews as mechanisms for spreading workflow technology knowledge. Specifically, finding ways to make proposal authors (typically domain scientists) aware of available technology so that their proposed work does not entail re-inventing the wheel (and thus spend more of their funding on the domain science) may be possible. Finally, it is clear that solving the ``community challenge'' has large overlap with solving the ``education challenge'' (Theme \#2.)
\end{itemize}

\medskip
\noindent
Breakout session participants identified the following {\bf short-term community effort}:

\begin{enumerate}
    \item \textbf{Establish a common knowledge-base for workflow technology}, for workflow users to be able to know and navigate the current technology landscape. User criteria (for navigation) need to be defined. Workflow system developers can self-report and add to this knowledge-based, and ideally, if workflow systems are deployed across sites, the knowledge base could include test status for a set of standard workflow configuration. There is large overlap with similar short-term community efforts identified in Themes~\#2 and~\#4.
\end{enumerate}

\medskip
\noindent
Breakout session participants also identified the following {\bf long-term community effort}:

\begin{enumerate}
    \item \textbf{Establish a ``Workflow Guild''}, i.e., an organization focused on interaction and good relationships and self-support between subscribing workflow developers and their systems, as well as dissemination of best-practices and tools that are used in the development and use of these systems.  Participants noted that:
    \begin{itemize}
        \item Such a community could be too self-reflecting, and yet still remain fragmented.
        \item A cultural/social problem is that creating a new system is typically more exciting for computer scientists as opposed to re-using someone's system.
        \item Building trust and reducing internal competition will be difficult, though building community identity will help the Guild work together against external competitors.
    \end{itemize}
\end{enumerate}


\newpage
\cleardoublepage\phantomsection\addcontentsline{toc}{section}{Appendix A: Participants and Contributors}
\section*{Appendix A: Participants and Contributors}
\label{appx:contributors}

\begin{longtable}[!h]{llp{10.5cm}}
\toprule
\textbf{First Name} & \textbf{Last Name} & \textbf{Affiliation} \\
\midrule
\rowcolor[HTML]{FFFFFF} 
Dong H.   & Ahn               & Lawrence Livermore National Laboratory                                                   \\
\rowcolor[HTML]{F2F2F2} 
Ilkay     & Altintas          & UC San Diego                                                                             \\
\rowcolor[HTML]{FFFFFF} 
Yadu      & Babuji            & University of Chicago                                                                    \\
\rowcolor[HTML]{F2F2F2} 
Rosa      & Badia             & Barcelona Supercomputing Center                                                          \\
\rowcolor[HTML]{FFFFFF} 
Vivien    & Bonazzi           & Deloitte                                                                                 \\
\rowcolor[HTML]{F2F2F2} 
Henri     & Casanova          & University of Hawaii at Manoa                                                            \\
\rowcolor[HTML]{FFFFFF} 
Kyle      & Chard             & University of Chicago                                                                    \\
\rowcolor[HTML]{F2F2F2} 
Tain\~a     & Coleman           & University of Southern California                                                        \\
\rowcolor[HTML]{FFFFFF} 
Michael   & Crusoe            & ELIXIR-NL; VU Amsterdam; Common Workflow Language project @ Software Freedom Conservancy \\
\rowcolor[HTML]{F2F2F2} 
Ewa       & Deelman           & University of Southern California                                                        \\
\rowcolor[HTML]{FFFFFF} 
Frank     & Di Natale         & Lawrence Livermore National Laboratory                                                   \\
\rowcolor[HTML]{F2F2F2} 
Paolo     & Di Tommaso        & Seqera Labs                                                                              \\
\rowcolor[HTML]{FFFFFF} 
Bjoern    & Enders            & NERSC                                                                                    \\
\rowcolor[HTML]{F2F2F2} 
Thomas    & Fahringer         & University of Innsbruck, Austria                                                         \\
\rowcolor[HTML]{FFFFFF} 
Rafael    & Ferreira da Silva & University of Southern California                                                        \\
\rowcolor[HTML]{F2F2F2} 
Rosa      & Filgueira         & JPMorgan Applied Research                                                                \\
\rowcolor[HTML]{FFFFFF} 
Grigori   & Fursin            & cTuning foundation and cKnowledge SAS                                                  \\
\rowcolor[HTML]{F2F2F2} 
Alex      & Ganose            & Lawrence Berkeley National Laboratory                                                    \\
\rowcolor[HTML]{FFFFFF} 
Carole    & Goble             & The University of Manchester                                                             \\
\rowcolor[HTML]{F2F2F2} 
Bj\"orn     & Gr\"uning           & Uni-Freiburg                                                                             \\
\rowcolor[HTML]{FFFFFF} 
Thuc      & Hoang             & Department of Energy                                                                     \\
\rowcolor[HTML]{F2F2F2} 
Shantenu  & Jha               & Rutgers University                                                                       \\
\rowcolor[HTML]{FFFFFF} 
Daniel S. & Katz              & University of Illinois at Urbana-Champaign                                               \\
\rowcolor[HTML]{F2F2F2} 
Olga      & Kuchar            & Oak Ridge National Laboratory                                                            \\
\rowcolor[HTML]{FFFFFF} 
Ana       & Kupresanin        & Lawrence Livermore National Laboratory                                                   \\
\rowcolor[HTML]{F2F2F2} 
Dan       & Laney             & Lawrence Livermore National Laboratory                                                   \\
\rowcolor[HTML]{FFFFFF} 
Bertram   & Ludascher         & University of Illinois at Urbana-Champaign                                               \\
\rowcolor[HTML]{F2F2F2} 
Ketan     & Maheshwari        & Oak Ridge National Laboratory                                                            \\
\rowcolor[HTML]{FFFFFF} 
Marta     & Mattoso           & Federal University of Rio de Janeiro                                                     \\
\rowcolor[HTML]{F2F2F2} 
Kshitij   & Mehta             & Oak Ridge National Laboratory                                                            \\
\rowcolor[HTML]{FFFFFF} 
Todd      & Munson            & Argonne National Laboratory                                                              \\
\rowcolor[HTML]{F2F2F2} 
Jonathan  & Ozik              & Argonne National Laboratory                                                              \\
\rowcolor[HTML]{FFFFFF} 
Tom       & Peterka           & Argonne National Laboratory                                                              \\
\rowcolor[HTML]{F2F2F2} 
Luc       & Peterson          & Lawrence Livermore National Laboratory                                                   \\
\rowcolor[HTML]{FFFFFF} 
Lo\"ic      & Pottier           & University of Southern California                                                        \\
\rowcolor[HTML]{F2F2F2} 
Lavanya   & Ramakrishnan      & Lawrence Berkeley National Laboratory                                                    \\
\rowcolor[HTML]{FFFFFF} 
Tim       & Randles           & Los Alamos National Lab                                                                  \\
\rowcolor[HTML]{F2F2F2} 
Stefan    & Robila            & National Science Foundation                                                              \\
\rowcolor[HTML]{FFFFFF} 
Stian     & Soiland-Reyes     & eScienceLab, The University of Manchester; BioExcel; Common Workflow Language            \\
\rowcolor[HTML]{F2F2F2} 
Alan      & Sussman           & National Science Foundation                                                              \\
\rowcolor[HTML]{FFFFFF} 
Douglas   & Thain             & University of Notre Dame                                                                 \\
\rowcolor[HTML]{F2F2F2} 
Benjamin  & Tovar             & University of Notre Dame                                                                 \\
\rowcolor[HTML]{FFFFFF} 
Matteo    & Turilli           & Rutgers University                                                                       \\
\rowcolor[HTML]{F2F2F2} 
Thomas    & Uram              & Argonne Leadership Computing Facility                                                    \\
\rowcolor[HTML]{FFFFFF} 
Karan     & Vahi             & University of Southern California
                            \\
\rowcolor[HTML]{F2F2F2} 
Michael   & Wilde             & Parallel Works Inc. and The University of Chicago                                        \\
\rowcolor[HTML]{FFFFFF} 
Matthew   & Wolf              & Oak Ridge National Laboratory                                                            \\
\rowcolor[HTML]{F2F2F2} 
Justin    & Wozniak           & Argonne National Laboratory                                                              \\
\bottomrule                                       
\end{longtable}

\newpage
\cleardoublepage\phantomsection\addcontentsline{toc}{section}{Appendix B: Agenda}
\section*{Appendix B: Agenda}
\label{appx:agenda}

\begin{table}[!h]
    \begin{tabular}{lp{12cm}}
        \toprule
        \textbf{Time} & \textbf{Topic} \\
        \midrule
        9:00-9:05am PST & \makecell[l]{\textbf{Welcome} and introductions\\\emph{\small Rafael Ferreira da Silva}} \\
        \rowcolor[HTML]{EEEEEE}
        9:05-9:15am PST & \makecell[l]{\textbf{workflowsRI} project overview\\\emph{\small Rafael Ferreira da Silva}} \\
        9:15-9:25am PST & \makecell[l]{\textbf{ExaWorks} project overview\\\emph{\small Dan Laney}} \\
        \rowcolor[HTML]{EEEEEE}
        9:25-9:55am PST & \textbf{Lightning talks}
            \begin{compactitem}
                \item Theme 1: FAIR computational workflows
                \item[] \emph{\small (Carole Goble)}
                \item Theme 2: Training and education for workflow users
                \item[] \emph{\small (Lavanya Ramakrishnan)}
                \item Theme 3: AI workflows
                \item[] \emph{\small (Luc Peterson)}
                \item Theme 4: Exascale challenges and beyond
                \item[] \emph{\small (Bjoern Enders)}
                \item Theme 5: APIs, interoperability, reuse, and standards
                \item[] \emph{\small (Douglas Thain)}
                \item Theme 6: Building a workflows community
                \item[] \emph{\small (Ilkay Altintas)}
                \vspace{-10pt}
            \end{compactitem} \\
        \rowcolor[HTML]{F7F7F7}
        9:55-10:05am PST & \emph{10min Break (preparation for breakout sessions)} \\
        \rowcolor[HTML]{EEEEEE}
        10:05-10:35am PST & \textbf{Breakout Session 1}
        \begin{compactitem}
                \item Theme 1: FAIR computational workflows
                \item Theme 2: Training and education for workflow users
                \item Theme 3: AI workflows
                \vspace{-10pt}
            \end{compactitem} \\
        10:35-10:50am PST & \textbf{Reports: Breakout Session 1} \\
        \rowcolor[HTML]{F7F7F7}
        10:50-11:05am PST & \emph{15min Break} \\
        \rowcolor[HTML]{EEEEEE}
        11:05-11:35am PST & \textbf{Breakout Session 2} 
        \begin{compactitem}
                \item Theme 4: Exascale challenges and beyond
                \item Theme 5: APIs, interoperability, reuse, and standards
                \item Theme 6: Building a workflows community
                \vspace{-10pt}
            \end{compactitem} \\
        11:35-11:50am PST & \textbf{Reports: Breakout Session 2} \\
        \rowcolor[HTML]{EEEEEE}
        11:50am-noon PST & \makecell[l]{\textbf{Final remarks} and moving forward\\\emph{\small Kyle Chard}} \\
        \bottomrule
    \end{tabular}
\end{table}

\end{document}